\documentclass[twocolumn,nofootinbib]{revtex4}

\makeindex

\usepackage{graphicx}
\usepackage{dcolumn}
\usepackage{bm}

\usepackage{amsmath}
\usepackage{amssymb}

\usepackage{etoolbox}

\usepackage[breaklinks=true,colorlinks,bookmarks=false,citecolor=blue,linkcolor=red]{hyperref}

\begin{document}


\title{A Machine Learning Inversion Scheme for Determining Interaction from Scattering}

\author{Ming-Ching Chang$^1$}
 
\author{Chi-Huan Tung$^{2,3}$}
\author{Shou-Yi Chang$^2$}
\author{Jan-Michael Carrillo$^4$}
\author{Bobby G. Sumpter$^4$}
\author{Changwoo Do$^3$}
\author{Wei-Ren Chen$^3$}
 
\affiliation{$^1$Department of Computer Science, University at Albany - State University of New York, Albany, New York 12222, United States}
\affiliation{ 
$^2$Department of Materials Science and Engineering, National Tsing Hua University, Hsinchu 300044, Taiwan}
\affiliation{
$^3$Neutron Scattering Division, Oak Ridge National Laboratory, Oak Ridge, Tennessee 37831, United States}
\affiliation{
$^4$Center for Nanophase Materials Sciences, Oak Ridge National Laboratory, Oak Ridge, Tennessee 37831, United States}


\begin{abstract}

We outline a machine learning strategy for determining the effective interaction in the condensed phases of matter using scattering. Via a case study of colloidal suspensions, we showed that the effective potential can be probabilistically inferred from the scattering spectra without any restriction imposed by model assumptions. Comparisons to existing parametric approaches demonstrate the superior performance of this method in accuracy, efficiency, and applicability. This method can effectively enable quantification of interaction in highly correlated systems using scattering and diffraction experiments.

\end{abstract}


\maketitle


Measuring interaction between particles in condensed matter has been of paramount interest since it provides a starting point for describing the statistical properties of the system under consideration. Elastic scattering and diffraction techniques have played an important role in this continued endeavor: From the measured spectra, extensive effort has been devoted to inferring the nature of forces that govern the properties of a variety of highly correlated disordered systems including dense atomic liquids~\cite{Schommers:1983,Aers:1984,Weis:1985,Laaksonen:1995}, ionic liquids~\cite{March:1984} such as electrolytes~\cite{Laaksonen:1999} and molten salts~\cite{Babu:1994,Caccamo:1996}, molecular fluids~\cite{Soper:1996,Soper:1999}, suspensions of colloids, micelles and emulsions~\cite{Macroions:Solution:1992,Nagele:1996,Likos:2001}, solutions of polymers~\cite{Schweizer:1992,Schweizer:1994}, and polyelectrolytes~\cite{Yethiraj:1996,Pollack:2006}. Nonetheless, potential inversion based on this experimental protocol is often hampered by the difficulty of precisely modeling the two-point static correlation functions, the quantity of interest in radiation scattering experiments, in terms of the relevant parameter. 

In this report we present a non-parametric strategy to circumvent the intrinsic limitation of existing approaches and demonstrate its feasibility by a case study of charged colloidal suspensions, a representative soft matter system. When the long-range electrostatic repulsion dominates over all distances, the effective interaction between charged colloids can be described by a hard sphere with a Yukawa tail of screened Coulomb repulsion~\cite{Pusey:1989}:
\begin{equation}
\beta \; V_{HSY}(r) = 
\begin{cases}
\infty, 
& \text{if } r < D\\
A \;
\frac{\exp\left[ -\kappa(r-D) \right]}
{r}, 
& \text{otherwise}
\end{cases}
\label{eq:Yukawa}
\end{equation}
where $\beta \equiv \frac{1}{k_B T}$ is the Boltzmann factor, $A$ the coupling parameter defined as $\frac{Z^2e^2}{\epsilon (1+ \frac{\kappa D}{2} )^2}$,  $D$ the colloidal diameter, $Z$ the charge number, $e$ the electric charge, $\epsilon$ the dielectric constant of solvent, and $\kappa$ the Debye screening constant. Eq.~\eqref{eq:Yukawa} has been extensively used to model the electrostatic interaction in a wide variety of charged colloidal systems including ionized nanoparticles~\cite{Macroions:Solution:1992}, self-assemblies~\cite{Zemb:1985}, and biological systems~\cite{Tardieu:1999}. The relevant correlation function is the inter-particle structure factor $S(Q)$ in reciprocal $Q$ space. One well-adopted approach to determine $V_{HSY} (r)$ from the measured $S(Q)$ is through the Ornstein-Zernike (OZ) integral equation~\cite{Hansen:1986,Macroions2:Solution:1992,Nagele:2004}: 
\begin{equation}
h(r_{12}) = c (r_{12}) + n_{p} \int c(r_{13}) h(r_{23}) d^{3} r_{3},
\label{eq:OZ}
\end{equation}
where $h (r_{12}) \equiv g (r_{12})-1$ and $g(r)$ is the pair distribution function, $c(r_{12})$ is the direct correlation function, and $n_{p}$ is the particle number density. $S(Q)$ is the Fourier transform of $g(r)$. Since both $h(r)$ and $c(r)$ are unknown, a second closure equation is required to solve Eq.~\eqref{eq:OZ}. Several closures have been developed as approximations~\cite{Macroions2:Solution:1992,Klein:DAguanno:1996,Nagele:1996}. Despite the popularity of the integral equation approach, existing studies have indicated its limitations: Because the accuracy of a given closure is not known $a$ $priori$, its validity in any specific phase region needs to be justified computationally~\cite{Belloni:1983,BeresfordSmith:1985,Belloni:1985,Belloni:1991,Fritz:2000,Banchio:2008,Heinen:2011,Heinen2:2011}. Moreover, the convergence behavior of each closure is found to depend on the complexity of adopted numerical procedures~\cite{Hus:2013}. 
For highly charged systems, the extraction of $V_{HSY} (r)$ from the corresponding $S(Q)$ is also known to be compounded by the strong electrostatic interactions~\cite{Anta:2002}. As alluded to above, no ideal closure is currently available. 

The position we take here is that developing another closure does not necessarily provide the most effective solution. From a Bayesian perspective~\cite{Murphy:ML:2012} we instead sought to solve this inversion problem via a machine learning (ML) approach based on Gaussian process~\cite{GPR:MIT:2006}, which defines a distribution over functions as a conceptual extension of the familiar Gaussian distribution. Using Eq.~\eqref{eq:Yukawa} we computationally generated an extensive library of $S(Q)$ from the equilibrium fluid phase \cite{SI}, which is defined as the training set $\{S_{train}(Q)\}$ in this study. In the vector space of $S(Q)$, we treated the $A$, $\kappa$ and $n_{p}$ of $\{S_{train}(Q)\}$ as a collection of normally distributed random variables and accordingly determined the statistical relationship of each variable in the ML process \cite{SI}. Using the optimized covariance matrix designed to quantify data correlation, we were able to probabilistically infer the values of $A$, $\kappa$ and $n_{p}$ from a given $S(Q)$ without having to rely on a prescribed parametric equation, such as OZ equation and a complementary closure, to specify this mathematical relationship deterministically. The uncertainties associated with the related parameters naturally emerge during the inference process embedded in the spectral analysis procedure.

\begin{figure}[t]
\centerline{
  \includegraphics[width=\linewidth]{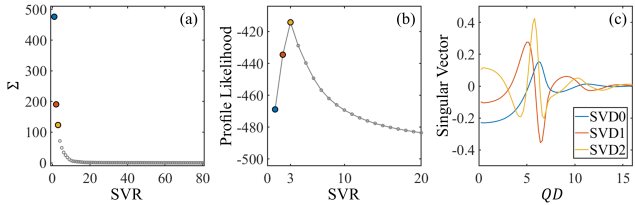}
  \vspace{-0.4cm}
}  
\caption{(a) The eigenvalues of principal components. As demonstrated in panel (b), the results of maximum profile likelihood analysis~\cite{Zhu:2006} show that the linear subspace spanned by the first three singular vectors given in Panel (c) is sufficiently expressive to illustrate the features of the correlated data points.           
}  
\label{fig:Scree}
  \vspace{-0.5cm}
\end{figure}

Before implementing the ML process, it is critical to first examine the feasibility of our proposed approach. To uniquely determine the potential parameters from a given $S(Q)$, a necessary condition to meet is the separability of dataset in the vector space of $S(Q)$ of dimension 80, the sampled $Q$ points in our computational trajectory analysis. Here we used a principal component analysis (PCA) by the singular value decomposition (SVD)~\cite{SVD:2016} to extract relevant information of the data distribution in this high dimensional vector space~\cite{SI}. From the results of singular value analysis~\cite{Zhu:2006} presented in Figs.~\ref{fig:Scree}(a) and ~\ref{fig:Scree}(b), it is confirmed that the variance of original data is mostly retained by the first three statistically significant singular value ranks. The vector space $\mathbb{R}^3$ spanned by these three singular vectors, denoted as SVD0, SVD1 and SVD2 in Fig.~\ref{fig:Scree}(c), is therefore sufficiently expressive in capturing essential features of the correlated data.

\begin{figure}[t]
\centerline{
  \includegraphics[width=\linewidth]{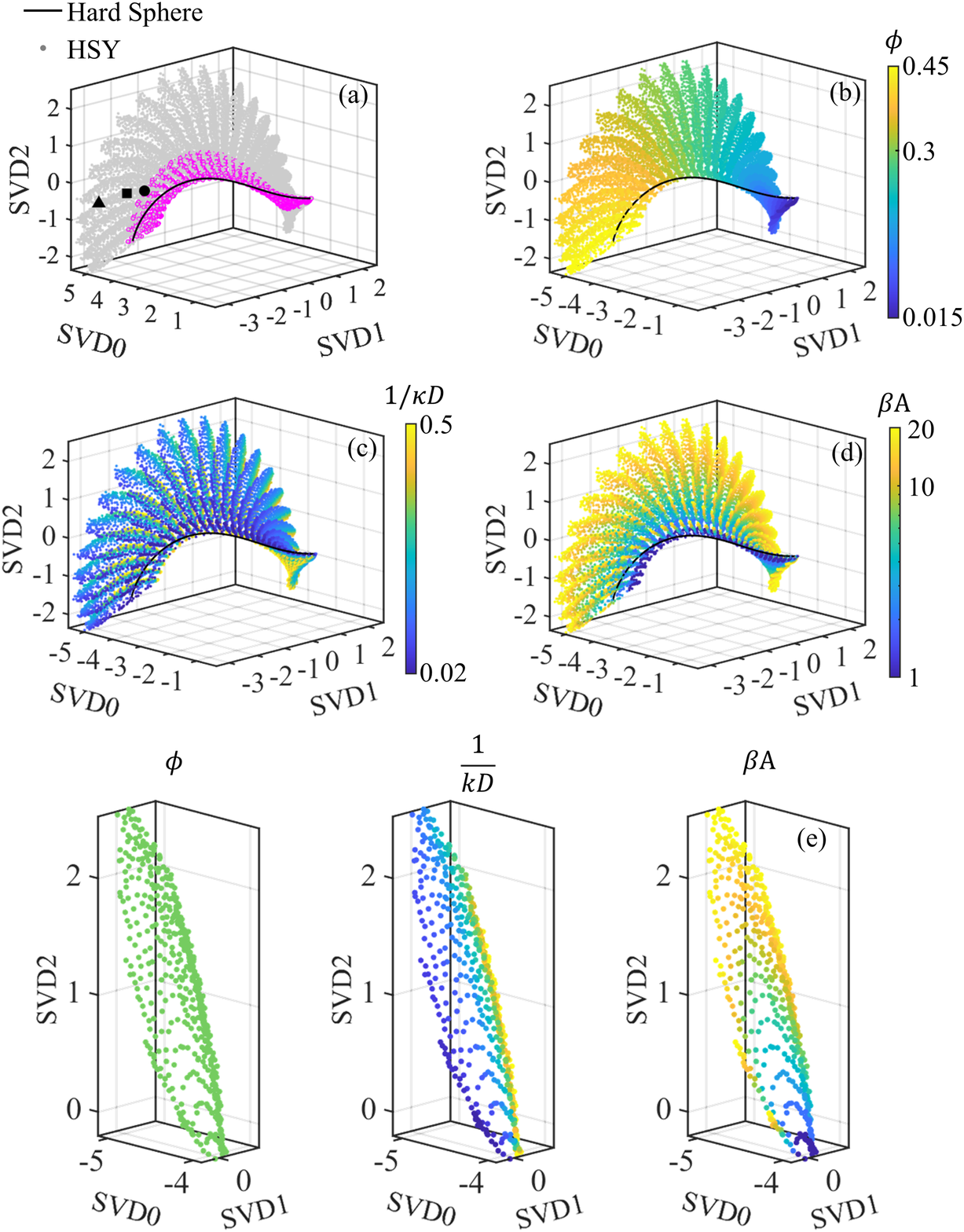}
  \vspace{0.0cm}
}  
\caption{
(a) The distributions of training data in the $\mathbb{R}^3$ vector space spanned by SVD0, SVD1 and SVD2 given in Fig.~\ref{fig:Scree}(c). The magenta symbols represent the fluid phase regions where the effective interaction can be precisely determined by solving the OZ equation with MPB-RMSA closure ~\cite{Heinen:2011,Heinen2:2011}. The blacks symbols represent the three highly correlated charged colloidal suspensions given in Fig.~\ref{fig:IQ}. Panels (b)-(d) give the distributions of $\phi$, $\frac{1}{\kappa D}$ and $\beta A$. The detailed distributions for $\phi=0.3$ are given as an example in (e).
}  
\label{fig:Fern}
\vspace{-0.4cm}
\end{figure}

In this $\mathbb{R}^3$ vector space, each $S(Q)$ in $\{S_{train}(Q)\}$ is represented by a point and the distribution of a gridded dataset is given in Fig.~\ref{fig:Fern}. As demonstrated by Fig.~\ref{fig:Fern}(a), it is seen to be narrowly distributed along one-dimension of a twisted three-dimensional manifold which resembles the shape of a half fern leaf: The elongated central axis represents the data of  of hard-sphere fluids with different volume fraction $\phi$ defined as $\frac{\pi}{6} n_{p} D^3$. Each pinna bursting forth from the central stalk consists of data points with same $\phi$ but different $\kappa$ and $A$. We thoroughly examined the data distribution presented in Fig.~\ref{fig:Fern}(a) and no inseparable overlapping was found. This observation, which reflects the one-to-one mapping between $S(Q)$ and a set of $\phi$, $\frac{1}{\kappa D}$, and $\beta A$ in colloidal fluid phase, provides initial support for a viable framework capable of inversely extracting potential parameters from measured scattering functions. The data points collected from the fluid phase characterized by $A$ $<$ 3 $k_{B}T$ are further marked by magenta color in Fig.~\ref{fig:Fern}(a). In this phase region the effective interaction of charged colloidal suspensions can be precisely determined by OZ equation complemented by the MPB-RMSA closure proposed by Heinen and coworkers~\cite{Heinen:2011,Heinen2:2011}, a sophisticated approach among the continued efforts~\cite{Hayter:1982,Hayter:1992} aiming at improving the MSA closure~\cite{Hayter:1981}. Within probed phase regions, its quantitative accuracy is found to be equivalent to that of Rogers-Young closure~\cite{RY:1984} but the computational efficient is significantly improved. One can further label the data points presented in Fig.~\ref{fig:Fern}(a) with the numerical values of $\phi$, $\frac{1}{\kappa D}$, and $\beta A$ to examine the characteristics of their distributions. Judging from the results given in Figs.~\ref{fig:Fern}(b)-\ref{fig:Fern}(d), the distributions of $\phi$, $\frac{1}{\kappa D}$ and $\beta A$ are all seen to vary smoothly. This observation suggests that the data points of these three parameters not only are self-avoiding but also highly correlated with a certain length scale. One can therefore uniquely extract the potential parameters from a given $S(Q)$ in its vector space. 

Having verified the feasibility of our proposed approach for spectral inversion, we can now readily demonstrate its numerical reliability. For this purpose it is instructive to further investigate the difference in heterogeneity of distribution for $\phi$, $\frac{1}{\kappa D}$, and $\beta A$. We found that the distribution of $\phi$ is a more slowly varying function in comparison to that of $\frac{1}{\kappa D}$ and $\beta A$. As exemplified by Fig.~\ref{fig:Fern}(e), for the pinna of $\phi=0.3$, both $\frac{1}{\kappa D}$ and $\beta A$ are seen to vary characteristically within $|x-x'|\sim{1}$ and $|x-x'|\sim{2}$ respectively, where $x$ and $x'$ are the coordinates in the vector space of $S(Q)$. In addition, the corresponding gradient vectors of $\frac{1}{\kappa D}$ and $\beta A$ are seen to point perpendicularly across and axially along the blade respectively. Clearly the numerical accuracy of potential extraction depends on how well this observed distributional heterogeneity is addressed. On the basis of Gaussian process we developed a covariance matrix, as the beating heart of our non-parametric inversion method, to quantitatively describe the statistical relationships of $\phi$, $\frac{1}{\kappa D}$ and $\beta A$ in the vector space of $S(Q)$~\cite{SI}. From $\{S_{train}(Q)\}$, the ML process was carried out to determine the optimal correlation lengths characterizing the covariance matrix to describe the local topological features of each parameter in the manifold of $\mathbb{R}^{80}$. For an input $S_{i}(Q)$, its statistical relevant neighbors can be identified from $\{S_{train}(Q)\}$ by comparing the Euclidean distance to $S_{i}(Q)$ and the optimized correlation length of each parameter. Therefore the $\phi$, $\frac{1}{\kappa D}$ and $\beta A$ of $S_{i}(Q)$, which must follow the correlation built-in in the covariance matrix if the system is in the equilibrium fluid state, can be forecasted probabilistically from those of its specified neighbors. 

\begin{figure}[t]
\centerline{
  \includegraphics[width=\linewidth]{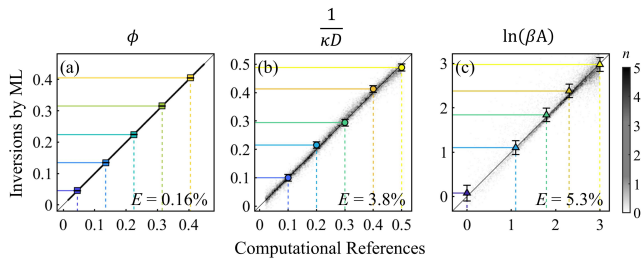}
  \vspace{-0.3cm}
}  
\caption{
The comparison of the extracted (a) $\phi$, (b) $\frac{1}{\kappa D}$, and (c) $\ln (\beta A)$ and their corresponding computational references. The numerical accuracy of reconstruction is quantified by the relative error $E$ given in each panel.               
}  
\label{fig:Fit}
\vspace{-0.6cm}
\end{figure}

Another extensive set of $S(Q)$ termed as $\{S_{test}(Q)\}$ was simulated separately from equilibrium fluid phase to gauge the numerical accuracy of our proposed method. Fig.~\ref{fig:Fit} presents the comparison of input parameters in simulations and those inverted from $\{S_{test}(Q)\}$. All three extracted parameters are in remarkable accord with their computational inputs, but $\phi$ is in closer quantitative agreement as indicated by the relative error $E$. The origin of this varying degree of uncertainty is worth exploiting: As illustrated in Fig.~\ref{fig:Fern}(e), both $\frac{1}{\kappa D}$ and $\beta A$ exhibit extreme changes within a relatively short Euclidean distance in comparison to the large-scale variation of $\phi$. This observation suggests that, the susceptibility of $S(Q)$ towards the variation of potential variables is inherently determined by their distributions in the vector space of $S(Q)$. The difference in $E$ can therefore be viewed a reflection of landscape heterogeneity for different parameters.  

\begin{figure}[t]
\centerline{
  \includegraphics[width=\linewidth]{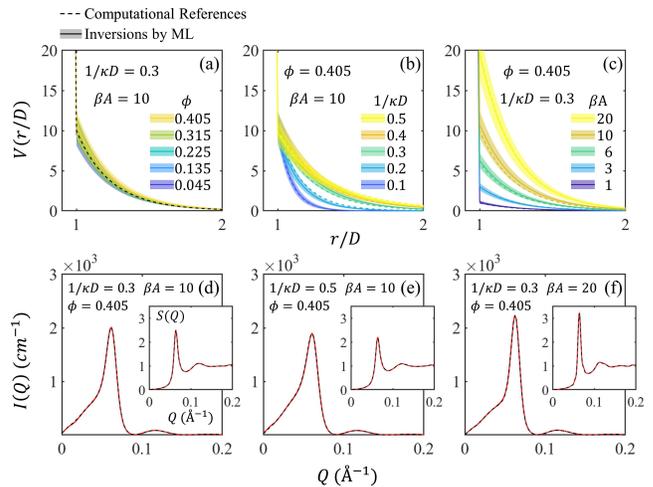}
  \vspace{-0.3cm}
}  
\caption{
Panels (a)-(c) present examples of $V_{HSY} (r)$ calculated based on $\phi$, $\kappa$ and $A$ given Fig.~\ref{fig:Fit}. The references used in computer simulations and those determined by ML inversion are presented in dashed and solid curves respectively. The magnitude of uncertainties calculated based on the errors in Fig.~\ref{fig:Fit} are represented by the shaded regions. The color scheme is same as that of Fig.~\ref{fig:Fit}. Panels (d)-(f) give the reconstructed $S(Q)$ (insets) and $I(Q)$ calculated from three highly correlated systems (yellow curves in panels (a)-(c) and upper right most points in Fig.~\ref{fig:Fit}). Both computationally generated reference $S(Q)$ and $I(Q)$ and their ML reconstructions are indistinguishable within the scales used in these plots. 
}  
\label{fig:IQ}
\vspace{-0.6cm}
\end{figure}

In Figs.~\ref{fig:IQ}(a)-\ref{fig:IQ}(c) we give the comparison of $V_{HSY} (r)$ calculated from the ML-extracted parameters specified by the solid lines in Fig.~\ref{fig:Fit} and their corresponding references used in simulation. To demonstrate the numerical accuracy of our approach, the $S(Q)$ of three highly correlated systems, as indicated by the black symbols in Fig.~\ref{fig:Fern}(a), were calculated and compared with their references in $\{S_{train}(Q)\}$. Accordingly the corresponding coherent intensities of small angle neutron scattering (SANS) $I(Q)$ were also calculated assuming the system is consisted of monodisperse densely-packed spherical particles~\cite{SI}. From the results presented in Figs.~\ref{fig:Fit}(d)-\ref{fig:Fit}(f), it is clearly seen that the differences between the reference $S(Q)$ (black dashed curves) and the reconstructed $S(Q)$ (red solid curves) are indeed indistinguishable on the scale of our plots. This observation demonstrates that the structural variation of suspending charged particles caused by the change of $V_{HSY} (r)$ with the magnitude of $E$ given in Figs.~\ref{fig:Fit}(a)-\ref{fig:Fit}(c) cannot be detected by elastic scattering techniques. The insets in Figs.~\ref{fig:IQ}(d)-\ref{fig:IQ}(f) also show that the difference in $I(Q)$ are within the range of statistical variation of resolution of general SANS instruments. For the three strongly correlated systems presented in Figs.~\ref{fig:IQ}(d)-\ref{fig:IQ}(f), integral equation theories are often unable to determine $\phi$ , $\kappa$ and $A$ from $S(Q)$ in a numerically precise manner~\cite{Anta:2002}. The validity of our proposed ML approach for quantitatively extracting $V_{HSY} (r)$ from the scattering spectra of charged colloidal suspensions over a wide range of $\phi$ in the equilibrium fluid phase is therefore verified. 

In conclusion, we have developed a new ML inversion method, based on the framework of Gaussian process, to inversely determine effective interaction of colloidal suspensions from their scattering spectra. By treating the probability distributions of the relevant potential parameters in the vector space of scattering function, our non-parametric approach circumvents the mathematical constraints inherent to deterministic models for spectral analysis. We demonstrated that our approach offers several advantages over the existing parametric approaches from the standpoint of numerical accuracy, computational efficiency~\cite{SI} and general applicability. Complemented by computer simulations, our method can be systematically extended for solving the inversion scattering problems of various colloidal systems characterized by different effective interactions~\cite{Asakura:1954,Witten:1986,YLiu:2005,Bianchi:2011}. Moreover, one recognized challenge in the analysis of scattering data of interacting systems is to take polydispersity into account adequately~\cite{Raj:1993}. In this pursuit, our ML strategy promises a new paradigm toward quantitative characterization of highly interacting systems characterized by significant variations of interaction potential and particle size which cannot be addressed precisely by existing decoupling approximations~\cite{Chen:1983, Hayter:1983,Pusey:1989}. 

In principle, the applicability of this method is not restricted to structural studies of colloidal systems. The essential idea underlying this method of spectral analysis, which does not suffer from the drawbacks of explicit modeling, allows quantitative extraction of relevant parameters based on which the targeted systems are computationally constructed. We are optimistic that our approach will provide a useful toolbox to facilitate the progress in many important inversion problems of radiation scattering, diffraction, and imaging~\cite{ConF1:1996,ConF2:1997,ConF3:2001,ConF4:2001} experiments from strongly correlated systems which traditionally have been difficult and time consuming to solve.

We thank Y. Shinohara, Y. Wang, P. Falus,  L. Porcar, Y. Liu, and S.-H. Chen for helpful discussions. This research was performed at he Spallation Neutron Source and the Center for Nanophase Materials Sciences, which are DOE Office of Science User Facilities operated by Oak Ridge National Laboratory. MD simulations used resources of the Oak Ridge Leadership Computing Facility, which is supported by DOE Office of Science under Contract DE-AC05-00OR22725. C.-H. T. thanks the financial support from the Shull Wollan Center during his stay at ORNL. BGS acknowledges support by the U.S. Department of Energy, Office of Science, Office of Basic Energy Sciences Data, Artificial Intelligence and Machine Learning at DOE Scientific User Facilities Program under Award Number 34532. M.-C. C. thanks the support provided by the University at Albany - SUNY. 

\bibliography{yukawaml}

\end{document}




\title{A Machine Learning Inversion Scheme for Determining Interaction from Scattering - Supplementary Material}

\author{Ming-Ching Chang$^1$}
 
\author{Chi-Huan Tung$^{2,3}$}
\author{Shou-Yi Chang$^2$}
\author{Jan-Michael Carrillo$^4$}
\author{Bobby G. Sumpter$^4$}
\author{Changwoo Do$^3$}
\author{Wei-Ren Chen$^3$}
 
\affiliation{$^1$Department of Computer Science, University at Albany - State University of New York, Albany, New York 12222, United States}
\affiliation{ 
$^2$Department of Materials Science and Engineering, National Tsing Hua University, Hsinchu 300044, Taiwan}
\affiliation{ 
$^3$Neutron Scattering Division, Oak Ridge National Laboratory, Oak Ridge, Tennessee 37831, United States}
\affiliation{
$^4$Center for Nanophase Materials Sciences, Oak Ridge National Laboratory, Oak Ridge, Tennessee 37831, United States}


\keywords{Colloidal interactions, hard-sphere Yukawa potential, Ornstein–Zernike (OZ) equation, closures, singular value decomposition, kernel ridge regression, Gaussian process, simulation. 
}

\maketitle











\section{Molecular Dynamics (MD) Simulation}
\label{sec:MD}

Using molecular dynamics (MD) simulation, over $30,000$ $S(QD)$ samples of charged colloidal suspensions in their equilibrium fluid phase were simulated based on Eq.(\textred{1}) in the main text. Experiments performed on $27,000$ samples of the dataset are reported in the following. The ranges of parameters are chosen to be 
$0 < \beta A < 20$, $0 < \phi < 0.45$, and 
$0 < \frac{1}{\kappa D} < 0.5$ respectively so that no splitting in the second peak~\cite{Glass:1995} of $S(QD)$, which signifies the formation of non-ergodic glassy state is observed. 
Based on the form of $\mu \equiv \{\phi, \kappa, A\}$ we separate the simulated $S(QD)$ into two main subsets: 

\begin{itemize}
\item \textbf{Grid15K} contains grid sample points with 30 unique $\phi$ values: $\phi \in (0.015, 0.03, 0.045, ..., 0.45)$, 25 unique $\kappa$ values: $\frac{1}{\kappa D} \in (0.02, 0.04, 0.06, ..., 0.5)$, and 20 unique $A$ values: $\beta A \in (1, 2, 3, ..., 20)$, such that there are totally $30 \times 25 \times 20 = 15,000$ sample points. It is the $\{S_{train}(Q)\}$ in the main text.
\item  \textbf{Rand12K} covers the specified $\mu$ range with $12,000$ uniform distributed random datapoints. It is the $\{S_{test}(Q)\}$ in the main text.
\end{itemize}

\begin{figure}[t]
    \centering
    \includegraphics[width=6cm]{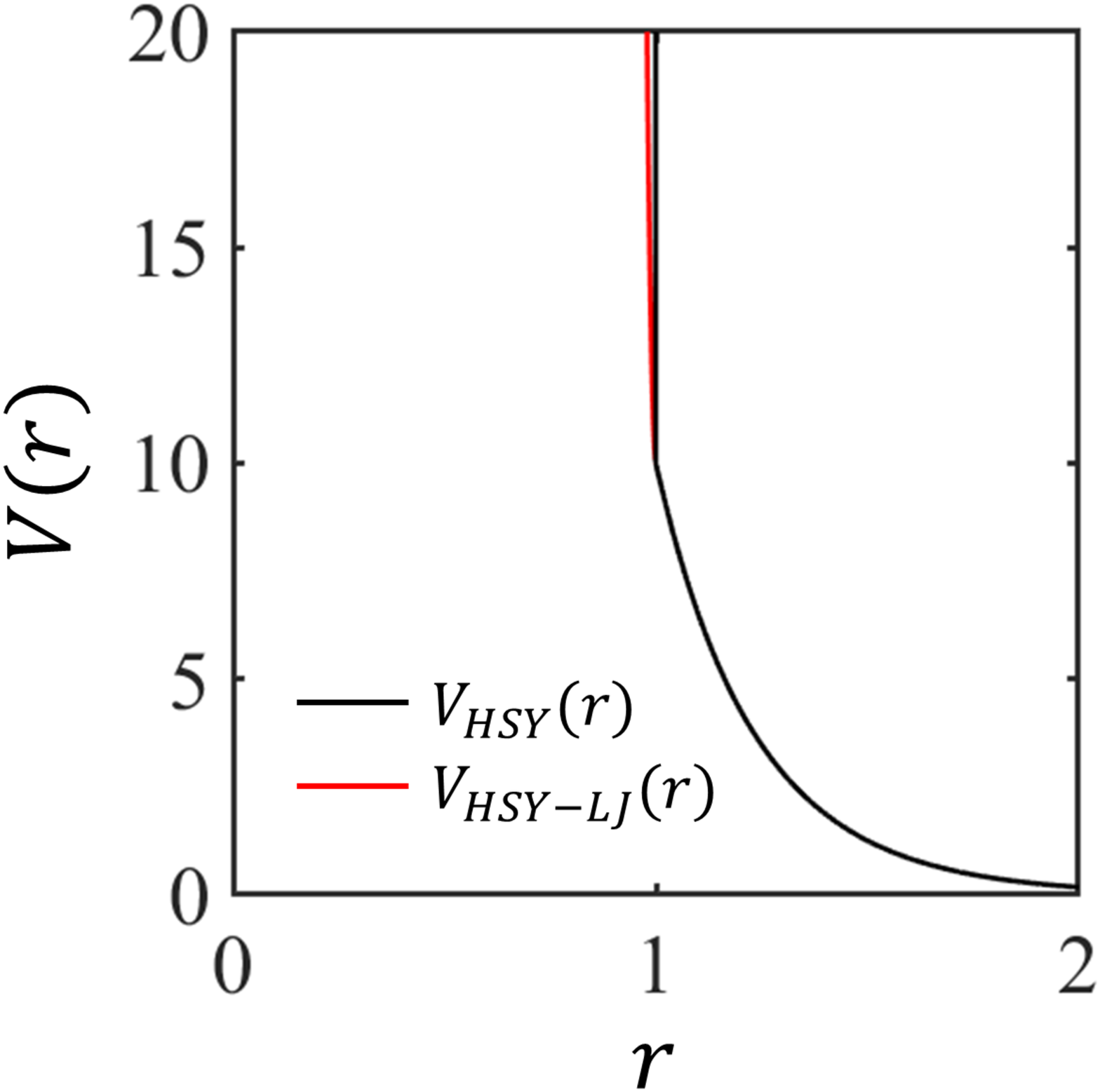}
    \vspace{-0.3cm}
    \caption{The hard-sphere Yukawa potential (black curve) and the approximated potential used in our simulations (red curve) by replacing the hard core with truncated-shifted Lennard-Jones interaction.
    }
    \label{fig:Vr_HSY_LJ}
\end{figure}
\begin{figure}[htp]
\centerline{
  \includegraphics[width=\linewidth]{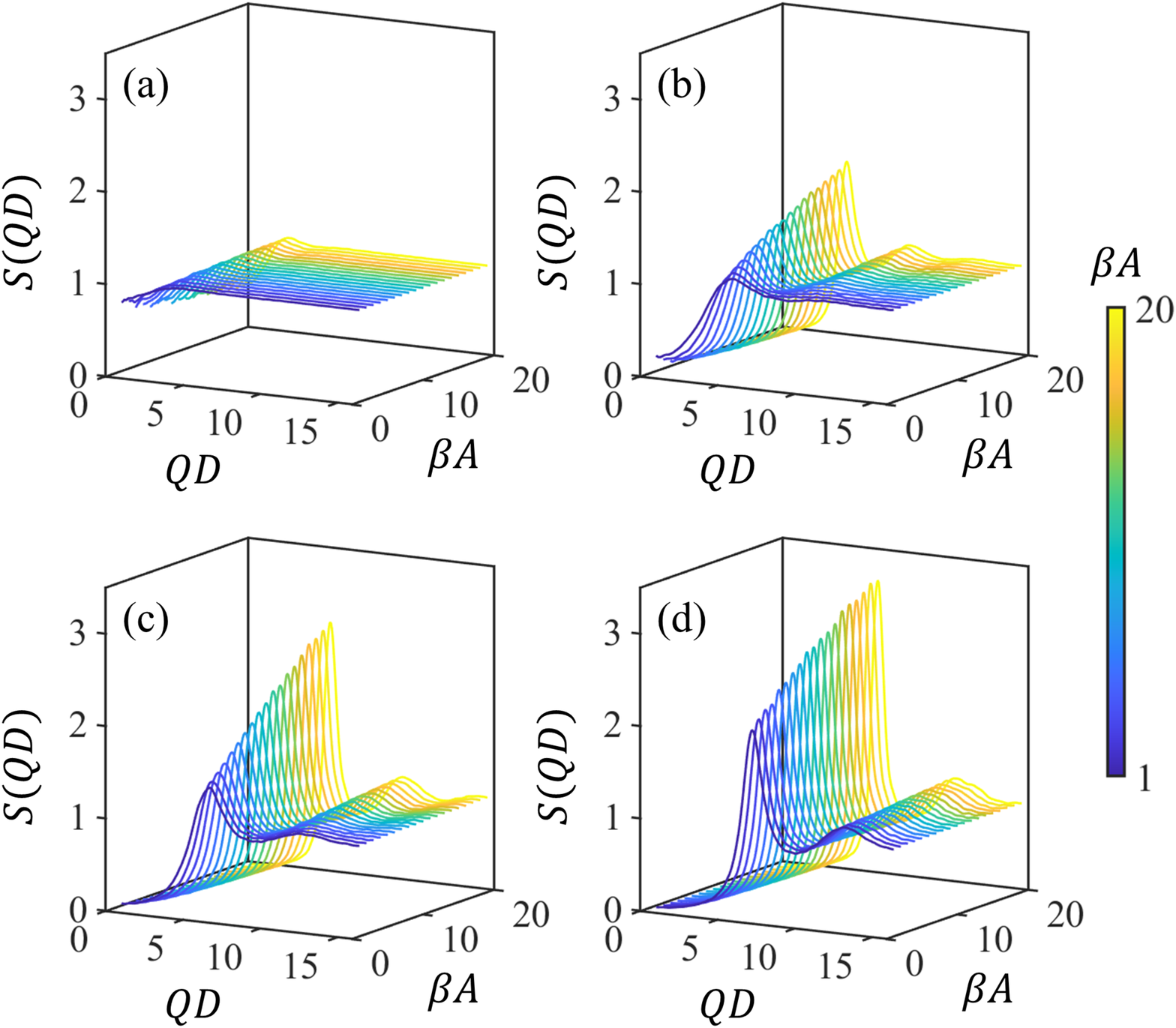}
  \vspace{-0.3cm}
}  
\caption{Examples of simulated $S(QD)$ with (a) $\phi = 0.015$, (b) $\phi = 0.15$, (c) $\phi = 0.30$, and (d) $\phi = 0.45$. The value of $\frac{1}{\kappa D}=0.3$ and $0 < \beta A < 20$.
}  
\label{fig:SQD}
\vspace{-0.6cm}
\end{figure}
The simulation cell contains $16,384$ particles with an initial number density of $n_{p} = \frac{3}{4\pi}\phi\  \delta^{-3}$. 
Canonical (NVT) ensemble simulation was performed, where the temperature was maintained at $T=1.0$ via a Nose-Hoover thermostat~\cite{Nose:1984,Hoover:1985} and integrated using the velocity-Verlet algorithm~\cite{Verlet:1967} with integration timestep of $0.001 \tau$. Here $\delta, \tau = \delta(\beta m)^{\frac{1}{2}}$, and $m=1$ are the standard LJ reduced units for distance, time, and mass respectively. 
The excluded hardcore of $D=1\ \delta$ in $V_{HSY} (r)$ used in our simulation is modeled by the repulsive component of 12-6 truncated-shifted Lennard-Jones (LJ) model~\cite{Hansen:1986}: 
\begin{equation}
V_{LJ}(r)= 
\begin{cases}
4\epsilon\left[(\frac{\sigma}{r})^{12}-(\frac{\sigma}{r})^{6}+\frac{1}{4}\right],
&\text{if}\ r<2^{\frac{1}{6}}\sigma\\
0,
&\text{if}\ r\geq 2^{\frac{1}{6}}\sigma\\
\end{cases}
\end{equation}
where $\beta\epsilon=500$ and $\sigma=2^{-\frac{1}{6}}\ \delta$ in reduced units. As displayed in Fig.~\ref{fig:Vr_HSY_LJ}, only marginal difference between the model potential $V_{HSY} (r)$ and $V_{HSY-LJ} (r)$ is observed. Each simulation proceeded for $5 \tau$ where the simulation box size is changed to specify a value of $\phi$, and another $25 \tau$ where the final $5 \tau$ was used to calculate $S(QD)$ using the existing procedure~\cite{carrillo2011polyelectrolytes,frigo1999fast}. Simulations were performed at the OLCF-4 Summit supercomputer of ORNL using the LAMMPS~\cite{Plimpton:1995} with graphical processing unit (GPU) acceleration. Examples of simulated $S(QD)$ are given in Fig.~\ref{fig:SQD}. Upon changing volume fraction $\phi=\frac{\pi}{6} n_{p} D^3$ and $A$, its evolution is consistent with documented results (for example see~\cite{Macroions:Solution:1992,Nagele:1996,Heinen:2011,Heinen2:2011}).
\section{SVD and PCA Analysis of MD Data}

As described in Section \ref{sec:MD}, we collected a dataset of $27,000$ $S(QD)$ of charged colloidal suspensions based on Eq.~(\textred{1}) of the main paper. 
We re-sampled each $S(QD)$ for $QD \in (0.2, 0.4, 0.6, ..., 16.0)$ such that $80$ $QD$ sample points in each simulated $S(QD)$ are kept. This way, the dimension of the vector space of $S(QD)$ is 80. Each $S(QD)$ is therefore represented by a point in this $\mathbb{R}^{80}$ vector space. Visualization of data distribution can be facilitated by dimensionality reduction: The data was arranged into a $80 \times 27,000$ matrix ${\bf F}$. Using Singular Value Decomposition (SVD)~\cite{SVD:2016}, ${\bf F}$ can be decomposed into 
${\bf F} = U \Sigma V^\mathsf{T}$.
The diagonal entries of $\Sigma^2$ are proportional to the percentages of the variance of the original data projected onto each corresponding principal axis. This Principal Component Analysis (PCA)~\cite{SVD:2016} allows us to re-express the data as a set of new orthogonal variables to extract their intrinsic correlations. To remove the subjectivity involved in deciding the correct number of component axes to retain, we use the analysis of maximum profile likelihood~\cite{Zhu:2006} to identify the statistical gap where the singular values begin to level off, as in Fig.~\textred{1}(a) of the main text. 
\section{Implementation of Machine Learning (ML)}
\begin{figure}[t]
\centerline{
  \includegraphics[width=\linewidth]{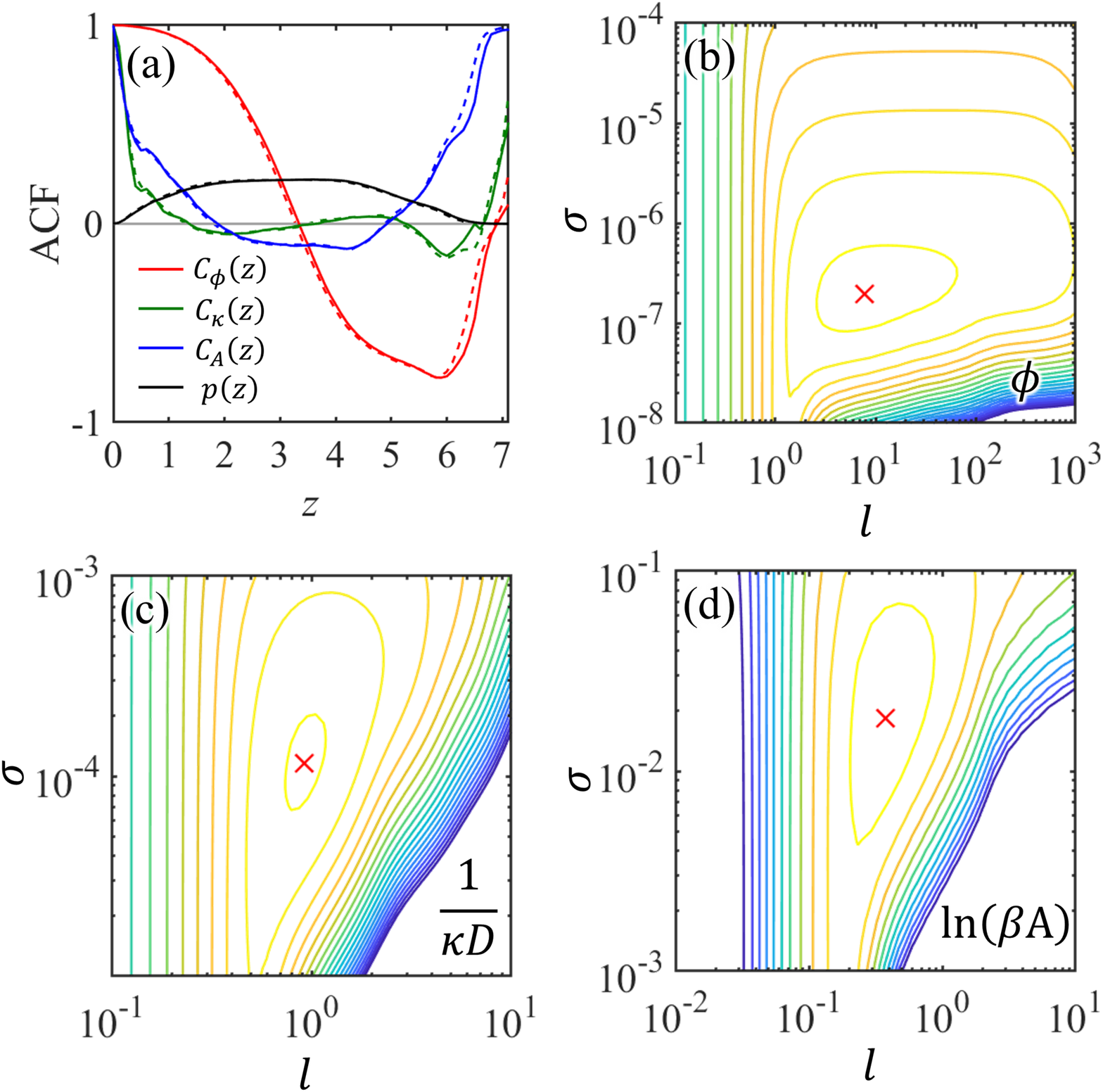}
  \vspace{-0.3cm}
}  
\caption{
The log marginal likelihood surfaces of $l$ and $\sigma$ for (a) $\phi$, (b) $\frac{1}{\kappa D}$, and (c) $\ln (\beta A)$. The optimal values are marked by the red crosses. Panel (d) gives the PDDF $p(z)$ of $\{S_{train}(Q)\}$ (black curve), and the auto correlations of $\phi$ ($C_\phi (z)$, red curve), $\kappa$ ($C_\kappa (z)$, green curve), and $A$ ($C_A (z)$, blue curve). Due to the sparsity of available pairs, the upturns observed in the range of $z>6$ lack statistical significance.               
}  
\label{fig:LML}
\vspace{-0.6cm}
\end{figure}
A function $f$ relating $S(QD)$ and $A$, $\kappa$ and $n_{p}$ can be formulated as $f \sim {\cal GP} (\mu, k)$ in terms of a prior mean function $\mu$ and a prior covariance function $k$ based on the framework Gaussian Process (GP)~\cite{GPR:MIT:2006}. Given s specific set $(X, Y)$, where $X$ represents the $n$ sets of $S(QD)$ in $\{S_{train}(Q)\}$ and $Y$ represents the corresponding $n$ sets of $A$, $\kappa$ and $n_{p}$, the purpose of the ML process is to determine $\mu$ and $k$ from the knowledge of the given data. Following the standard procedure of GP~\cite{GPR:MIT:2006}, a constant function is used to specify $\mu$. The $n \times n$ covariance matrix $K_{XX}$ specifies the correlations between training data pairs modeled by the radial basis function (RBF) kernel. Specifically, for $x, x' \in X$, kernels $k(x, x')$ as entries of $K_{XX}$ are formulated by the following squared exponential expression:
\begin{equation}
k(x, x') = \exp 
\left[
\frac{-(x-x')^2}{2 l^2}
\right],
\label{eq:rbf}
\end{equation}
where 
$l$ is the correlation length which measures the similarity between training data points. The Gaussian observational noise term is added to the kernel matrix $K = K_{XX} + \sigma^2 I$ where $\sigma$ describes the variance of observational noise. Given a test point $X_*$, the goal of ML is to estimate $Y_* = f(X_*)$, where the convariance matrix of any test set selected from $\{S_{test}(Q)\}$ is denoted as $K_{X_* X_*}$, and $K_{X X_*}$ consists of entries measuring the correlations between training and test points. 
For simplicity, $K \equiv K_{X}$, $K_* \equiv K_{X X_*}$, and $K_{**} \equiv K_{X_* X_*}$.  For a $S(QD)$ measured from the equilibrium fluid phase, the relevancy of its $A$, $\kappa$ and $n_{p}$ with those in $\{S_{train}(Q)\}$ should follow the same correlation patterns deduced from the training process. Therefore, they can be determined from this joint distribution:
\begin{equation}
\left[
\begin{array}{c}
Y \\
Y_*
\end{array}
\right]
\sim
{\cal N}
\left(
\left[
\begin{array}{c}
\mu(X) \\
\mu(X_*)
\end{array}
\right],
\left[
 \begin{array}{cc}
 K & K_*^\mathsf{T} \\
K_* & K_{**} \\
\end{array}
\right]
\right),
\end{equation}
where $\mathsf{T}$ indicates matrix transposition, $\cal N$ denotes normal distribution. The parameters in Eq.~\eqref{eq:rbf} is determined by maximizing the {\em log marginal likelihood} ${\cal L}$ using gradient descent algorithm during training~\cite{GPR:MIT:2006,Exact:GP:NIPS2019}. 
To facilitate the optimization process, we first investigate the packing pattern of data in the manifold using the pair distance distribution function (PDDF)~\cite{glatter1979interpretation}:
\begin{equation}
p(z)
=
\frac{1}{N^2}
\sum_{i=1}^{N}
\sum_{j=1}^{N}
\delta \left(
|x_{i}-x_{j}| - z
\right),
\label{eq:PDDF}
\end{equation}
where $\delta$ is the Dirac delta function. The shape of PDDF (black curve) given in Fig.~\ref{fig:LML}(a) suggests a nearly evenly distribution of data points~\cite{glatter1979interpretation}. 

Because no heterogeneous clustering of data is observed, the expectation values of the potential parameters associated with any given data point can be determined by its correlation between other data points through the optimized kernel functions used in GP. Information about the distributions of $\phi$ , $\frac{1}{\kappa D}$ and $\beta A$ can be further deduced from their autocorrelation functions (ACF)~\cite{debye1957scattering}:
\begin{equation}
C_{\mu}(z)
=
\frac{
\left<
\mu(x)
\mu(x+z)
\right>_x
-\left<\mu(x)\right>_{x}^{2}
}
{\left<
[\mu(x)]^2
\right>_x
-\left<\mu(x)\right>_{x}^{2}
},
\label{eq:autocorr}
\end{equation}
where $\left< \cdot \right>_x$ represents the spatial average over position $x$ of all sample points in the vector space. From the results given in Fig.~\ref{fig:LML}(a) $C_\phi (z)$ is characterized by a slow relaxation with a characteristic length which is larger than the scale of the probed spatial domain. This observed decaying behavior is consistent with the smooth variation of $\phi$ revealed by Fig.~\textred{2}(b) in the main text. On contrary, the self-correlations of $\kappa$ and $A$ are no longer retained when $z>2$. This result gives a quantitative measure regarding the average length of their distributions shown in Fig.~\textred{2}(e) in the main text. Because the length scale estimated by Eq.~\eqref{eq:autocorr} is a spatial range beyond which each parameter is allowed to vary randomly, it therefore defines the upper limit of the correlated spatial range of each parameter. 
Figs.~\ref{fig:LML}(b)-\ref{fig:LML}(d) presents the marginal likelihood surfaces over the two relevant parameters, the correlation length $l$ and the variance of observational noise $\sigma^2$, the central ingredients of the covariance matrix describing the correlation of training data on the framework of GP. 
For $\phi$ , $\frac{1}{\kappa D}$ and $\ln (\beta A)$, the contours are seen to be unimodal and convex with different degrees of skewness which reflects the characteristic patterns of their distributions in the vector space of $S(Q)$. The determination of the optimal $l$ and $\sigma^2$ for $\phi$ , $\frac{1}{\kappa D}$ and $\beta A$ is greatly facilitated by these monotonic features. The optimized $l$ and $\sigma^2$ obtained from gradient descent are marked by red cross symbols in Figs.~\ref{fig:LML}(b)-\ref{fig:LML}(d). It is instructive to indicate that the optimized $l$ for each parameter are indeed less than its correlation length determined by ACF analysis given in Fig.~\ref{fig:LML}(a). In this paper we adopt the \texttt{sklearn GaussianProcessRegressor} library~\cite{SK:2011} due to its efficiency and convenience of implementation. We give the results of optimization in Table~\ref{tab:hyperparameters}. It is found that $l_{\phi} = 7.74$, $l_{\kappa} = 0.922$, $l_{A} = 0.373$, and $\sigma_{\phi} = 1.94\times10^{-7}$, 
$\sigma_{\kappa} = 1.16\times10^{-4}$,
$\sigma_{A} = 1.83\times10^{-2}$,
respectively. The numerical accuracy of these values are cross-validated. It is also noted that they are consistent with the calculations of ACF given in Fig.~\ref{fig:LML}(a).

\begin{table}[htp]
\caption{$l$ and $\sigma$ obtained via gradient descent optimization.}
\centerline{
\begin{tabular}{|c|c|c|}
\hline
$\theta$ & $l$ & $\sigma$ \\
\hline\hline
$\phi$ & 7.74 & $1.94\times10^{-7}$ \\
\hline
$\frac{1}{\kappa D}$ & 0.922 & $1.16\times10^{-4}$ \\ 
\hline
$\ln(\beta A)$ & 0.373 & $1.83\times10^{-2}$ \\ 
\hline
\end{tabular}
}
\label{tab:hyperparameters}
\end{table}

\section{Coherent Scattering Cross Sections $I(Q)$ of Small Angle Neutron Scattering (SANS)}

\begin{figure}
    \centering
    \includegraphics[width=\linewidth]{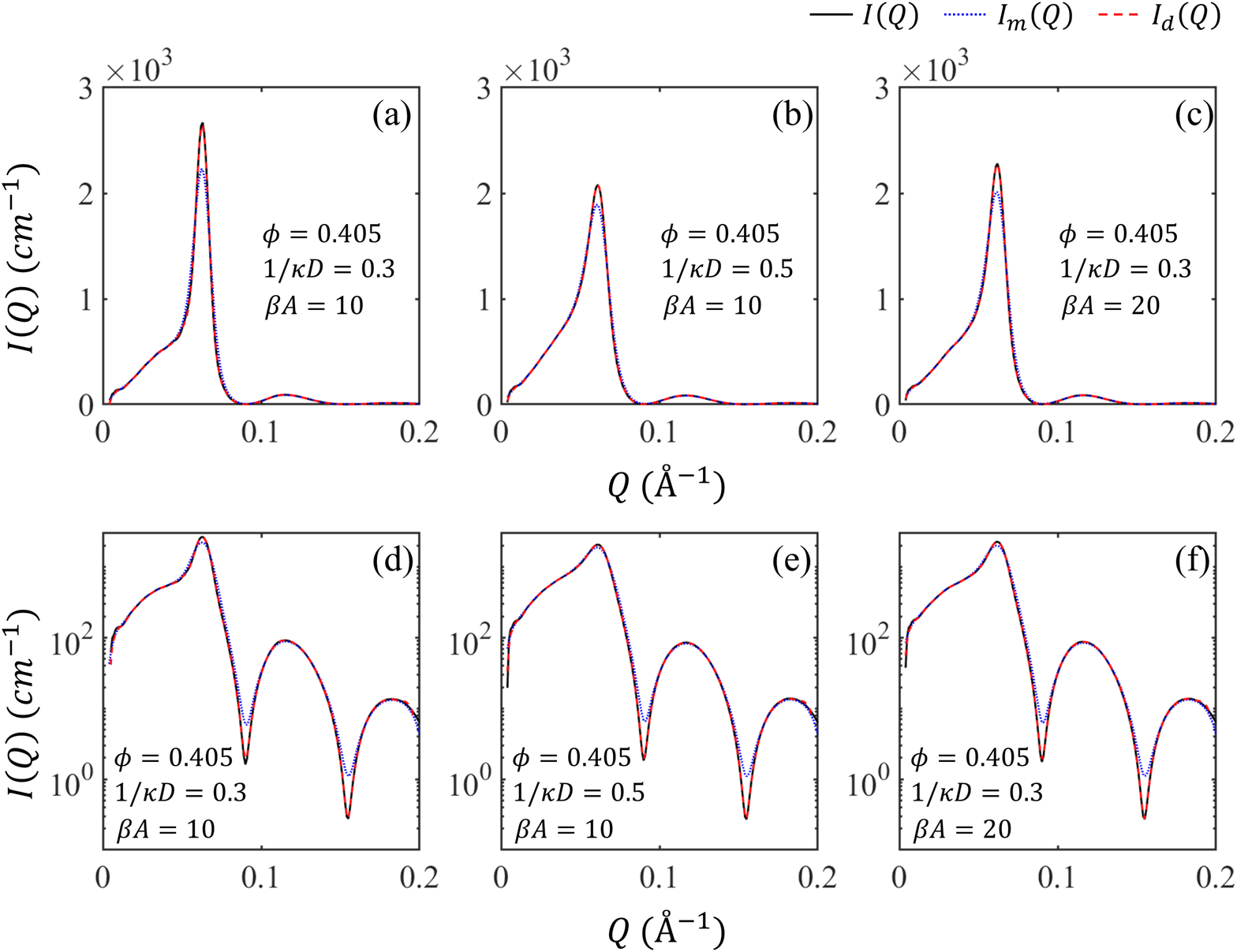}
    \vspace{-0.3cm}
    \caption{SANS coherent scattering intensities calculated based on the system of monodispersed charged $SiO_{2}$ particles immersed in $D_{2}O$. $I(Q)$ (black solid curves) represents the scattering intensity calculated from the computationally generated $S(Q)$ and $P(Q)$, $I_{m}(Q)$ (blue dotted curves) the intensity obtained by convolving $I(Q)$ with the instrument resolution, and $I_{d}(Q)$ (red dashed curves) the intensity obtained by removing the instrument resolution from $I_{m}(Q)$. Panels (a)-(c) display the data in linear scale and (d)-(f) are the same data presented in semi-log scale.
    }
    \label{fig:Lake}
\end{figure}

SANS $I(Q)$ for an interacting colloidal system consisting of monodisperse spheres can be expressed as~\cite{Chen:1986}:

\begin{equation}
I(Q) = n_{p}(\Delta\rho)^2(v)^2P(Q)S(Q),
\label{eq:CX}
\end{equation}
where $n_{p}$ is the number density of particle, $\Delta\rho$ the difference between the scattering length density of particle and that of solvent, $v$ the volume of particle, $P(Q)$ the form factor, and $S(Q)$ the inter-particle structure factor. In this work $P(Q)$ is calculated by the hard sphere model~\cite{Pedersen:1997}. The diameter is set to be $D$ the excluded hardcore of $V_{HSY} (r)$. The caculated scattering intensities are further convoluted with the instrument resolution function based on the resolution of EQ-SANS instrument at SNS ORNL. The selected system is silica particles dispersed in deuterium oxide. The density of deuterium oxide is 1.1 $g/ml$. The diameter and density of silica particle are set to be 100 {\AA} and 2.2 $g/ml$ respectively. Accordingly $\Delta\rho$ can be calculated~\cite{Chen:1997}. Examples of calculated $I(Q)$ are given in Fig.~\ref{fig:Lake}. It is worth noting that the effect of instrument resolution can be satisfactorily removed based on exiting protocols~\cite{Lake:2011}.

\section{Efficiency of Our ML Inversion Approach}

The computational demand of our method is extremely lightweight when compared to existing integral equation approaches. Our method does not require large-scale GPU cloud. The whole training processes can be completed in hours, and testing can be completed in minutes or seconds.
On a standard i7 laptop computer without GPU, our implementation of our ML process for $15,000$ data points takes about 3 hours to finish and a few minutes of for testing all the other $12,000$ data points. In general, solving an inversion problem from one $S(Q)$ curve takes less than $10^{-3}$ s. In comparison, it takes approximately 0.1 s to generate a $S(Q)$ from a specific combination of $\phi$, $\kappa$ and $A$ using the state-of-art integral equation approach~\cite{Heinen:2011,Heinen2:2011}. Moreover, for a general curve fitting routine, hundreds of iterations are often required to obtain satisfactory agreement between parametric model and experimental data. Overall the efficiency is improved significantly by our non-parametric ML approach in comparison to the existing parametric integral equation approaches.

\bibliography{yukawaml}
